# The benefit of law-making power

Anshul Gupta and Sven Schewe

University of Liverpool

**Abstract.** We study optimal equilibria in multi-player games. An equilibrium is *optimal* for a player, if her payoff is maximal. A tempting approach to solving this problem is to seek optimal Nash equilibria, the standard form of equilibria where no player has an incentive to deviate from her strategy. We argue that a player with the power to define an equilibrium is in a position, where she should not be interested in the symmetry of a Nash equilibrium, and ignore the question of whether or not her outcome can be improved if the other strategies are fixed. That is, she would only have to make sure that the other players have no incentive to deviate. This defines a greater class of equilibria, which may have better (and cannot have worse) optimal equilibria for the designated powerful player. We apply this strategy to concurrent bimatrix games and to turn based multi-player mean-payoff games. For the latter, we show that such political equilibria as well as Nash equilibria always exist, and provide simple examples where the political equilibrium is superior. We show that constructing political and Nash equilibria are NP-complete problems. We also show that, for a fixed number of players, the hardest part is to solve the underlying two-player mean-payoff games: using an MPG oracle, the problem is solvable in polynomial time. It is therefore in UP and CoUP, and can be solved in pseudo polynomial and expected subexponential time.

## 1 Introduction

Nash equilibria [17, 15, 22, 18, 8] are a common way to describe stable strategies with the intuition that only if no player gains from changing her strategy unilaterally, the strategy will be maintained. In this paper, we raise the question of how an interested party, called *dictator* in the remainder, can capitalise on setting an equilibrium strategy.

*How should a selfish agent design the rules if she has a chance to do so?*

In response to this, the policy of the dictator is to optimise her payoff by chosing a strategy she could improve upon. This conceptual contribution can quite simply be demonstrated by the following concurrent bimatrix games, which are extensions of the prisoners dilemma. Prisoner I takes the role of the dictator in our examples.

The first game is defined by the left bimatrix shown in Table 1. Both prisoners have the common options to co-operate with (C) or defect (D) the police. If both defect, they will be sentenced for minor crimes (receiving a one year sentence, reflected by a payoff of $-1$). If both co-operate, they receive an eight year sentence. If one prisoner co-operates while the other prisoner defects, the defector expects a ten year sentence, while the co-operator will receive a key witness status and won't be charged. Prisoner I, however, has a third option, where she can play politically (P) by commit to some of the crimes, but (with the help of her expensive lawyer) in a way that the charge is not a full one. In this case, Prisoner I will receive a charge of five years if Prisoner II defects, and a charge of nine years if she co-operates. Prisoner II will receive an eight year sentence either way (in case of defection as a beneficiary of the defence strategy of Prisoner I).

|  | Prisoner I |  |  |
|---|---|---|---|
| Prisoner II | D | C | P |
| D | -1 / -1 | 0 / -10 | -5 / -8 |
| C | -10 / 0 | -8 / -8 | -9 / -8 |

|  | Prisoner I |  |  |
|---|---|---|---|
| Prisoner II | D | C | P |
| D | -1 / -1 | 0 / -10 | -5 / -5 |
| C | -10 / 0 | -8 / -8 | -9 / -8 |

**Table 1.** Payoff matrices; the upper left / lower right values refer to Prioner I / II.

In this case, the only Nash equilibrium is (as in the classic prisoners dilemma) that both prisoners co-operate. The optimal *political* equilibrium for Prisoner I, however, is to play politically (P), while Prisoner II defects (D). This strategy yields a payoff of $(-5, -8)$ and is an (optimal) political equilibrium, because the *other* prisoner does not have any incentive to deviate from the strategy assigned to her by the dictator, but it is not a Nash equilibrium, as Prisoner I has an incentive to deviate, both to C and to D. (For all other dictator strategies, co-operating is the only optimal counter strategy of Prisoner II.)

In the right matrix, the only difference is that Prisoner II benefits fully from the defence strategy of Prisoner I when Prisoner I plays P. In this case, Prisoner I has a mixed optimal equilibrium, namely to play D with a $\frac{3}{4}$, and $P$ with a $\frac{1}{4}$ probability, while assigning Prisoner II the pure strategy to defect. This optimal policital equilibrium yields a payoff of $(-2, -2)$. (Note that C/C remains the only Nash equilibrium, as Prisoner I will still, for all strategies of Prisoner II, benefit most from co-operating, and the best response of Prisoner II to a co-operating Prisoner I remains to play C).

**Application in mean-payoff games.** In the remainder, we focus on the application of political equilibria in turn based mean-payoff games. Mean-payoff games (MPGs) [24, 11, 9, 2, 19, 3, 6] are finite turn-based games of infinite duration. They are played on a game arena, a directed graph, whose vertices are owned by different players. An MPG is played by placing a token on a vertex, and allowing the player who owns the vertex to push the token forward along an edge in the arena. Thus, the players successively create an infinite play. The edges of the game hold rewards for each player, and the objective of each player is to maximise her average reward.

The way each individual player plays can be captured by a strategy, and a set of strategies, one for each player, is called a strategy profile. A strategy profile is a Nash equilibrium if no player has an incentive for unilateral deviation: if all other players stick to their strategy, she cannot increase her payoff by changing hers.

In mean-payoff games, it is simple to think of scenarios where the power of the players is not symmetric. Think, for example, of a client-server scenario, where the server provides a service and can therefore *dictate* the rules. One may also think of political processes, where rules are laid down in laws, bylaws, conditions of service, or just a social etiquette. It seems natural that players in the position to change these rules have a greater power over defining an equilibrium.

Apart from this natural motivation, it is always reasonable to ask which equilibrium is best, be it for an individual player or for society. Especially if we seek the optimal strategy of an individual player (the dictator), it does not seem natural to restrict her choice of a strategy any further than necessary.

An arguably necessary restriction is that no *other* player has an incentive to deviate from the strategy defined, just as it is wont in Nash equilibria. But if we allow the dictator to select strategies that she can improve over, we give her more leeway when selecting a strategy profile. We therefore argue that we should allow her to 'discriminate' against herself. In consequence, the dictator may assign strategy profiles in such a way that she could improve over the outcome provided that the other players stick to the strategy she has assigned to them.

As we will show in Section 3.1, the dictator may suffer from restricting her strategy in the Nash sense. In our opinion, there is no convincing reason why she should constraint her strategies in this way. Note that the society may be viewed as a player without moves, such that the interest of society can be viewed as a special case.

**Results.** The main contribution of this paper is the introduction of the concept of political equilibria. We believe that political equilibria are a natural question that arise when we seek to construct a stable strategy assignment.

If we start with the question of Nash equilibria, a natural follow-up question is *which* Nash equilibrium we should choose in scenarios where there are many. Choosing the optimal Nash equilibrium for one of the players seem to be a very natural question.

Once this question is asked, a very natural follow-up question is why we should restrict the moves of this designated player unnecessarily. When we give this player the defining power over the equilibrium, any restriction that aim at her interests is either void or harmful. The natural consequence is to lift all of these restrictions in order to provide her with as much leeway as possible.

For multi-player mean-payoff games (MMPGs), we provide additional contributions. We show that (unsurprisingly) the complexity of finding optimal political equilibria equals the complexity of finding Nash equilibria. We show that optimal equilibria always exists. The complexity of finding a given Nash equilibrium is known to be NP complete [23]. We show that NP completeness (unsurprisingly) extends to optimal equilibria, and sharpen this bound by showing that they cannot be approximised. We also show that hardness depends on the number of players: for a bounded number of players, we give a polynomial time reduction to solving 2 player mean-payoff games (2MPGs).

As the complexity of solving 2MPGs is wide open, we cannot hope for determining the precise complexity for solving MMPGs with a bounded number of players without establishing the complexity of solving 2MPGs. However, we get simple corollaries for the complexity of finding political and Nash equilibria for a bounded number of players: this can be done in pseudo polynomial time [6], this can be done in smoothed polynomial time [3], there are fast randomised [2] and deterministic [19] strategy improvement algorithms, and the decision problem is in UP∩CoUP [11, 24]. This reduction hinges on the fact that very simple strategies, to which we refer to as reward and punish strategies, suffice. Reward and punish strategies essentially dictate a play of the game. Upon deviation, the game turns into a two-player game (hence the reduction), where the player who deviates first will henceforth play against the remaining players.

Finally, we discuss how to use our results to find equilibria that are best for society rather than egoistic ones that suit only a single player.

**Related Work.** The existence of Nash equilibria has recently been established by Brihaye, Bruyère, and De Pril [4]. Their proof has been significantly simplified and the existence of very simple strategies has been established in [5]. Our reward and punish strategy profiles are inspired by the simple strategies used in [5] and similar strategies in stateless games [26].

The existence of Nash equilibria in multi-player mean-payoff games has been established in [25]. Ummels and Wojtczak [23] studied the complexity of determining the existence of Nash equilibria, where each reward falls into a given closed interval in multi-player mean-payoff games. Both sides of the NP completenss proofs are closely related to ours. In [21], Ummels has studied the concept of subgame perfect equilibrium in case of infinite games. He has given simple examples to show that subgame perfect equilibrium, where choice of strategy should be such that it is optimal for initial history of the game and not for just initial vertex, exists in the case of infinite games.

There are quite a few works on optimal equilibria, in particular on equilibria that are 'best for society', which is usually defined as the optimal sum. This definiton is, for example, used in the definition of the price of Anarchy [14] for network and internet related games, or in traffic routing games [1, 10]. In [16], the authors study a virus inoculation game on social networks, in which players think of their neighbour's welfare. In [7], the authors have modelled a society game and shown how the equilibria are affected if players think of society rather than thinking of themselves.

## 2 Preliminaries

A *multi-player mean-payoff game* (MMPG) is a tuple $\langle P, V, \{V_p \mid p \in P\}, v_0, E, \{r_p : E \to \mathbb{Q} \mid p \in P\}\rangle$, where $P$ is a set of players, $V$ is a set of vertices with a designated initial vertex $v_0 \in V$, $\{V_p \mid p \in P\}$ is a partition of the vertices $V$, $E \subseteq V \times V$ is a set of edges, such that each vertex has a successor ($\forall v \in V \ \exists v' \in V, \ (v, v') \in E$), and $\{r_p \mid p \in P\}$ is a family of reward functions $r_p : E \to \mathbb{Q}$, that assign, for each respective player $p$, a reward for each transition to $p$.

An MMPG is intuitively played by placing a token on the initial vertex. Each time the token is in the vertex of a player $p$, player $p$ chooses an outgoing transition and moves the token along this transition. This way, the players jointly construct an infinite *play* $\pi \in V^\omega$. For each player $p$, a play $\pi = v_0, v_1 \ldots$ is evaluated to

$$r_p(\pi) = \liminf_{n \to \infty} \frac{1}{n} \sum_{i=0}^{n-1} r_p\big((v_i, v_{i+1})\big).$$

If the reward functions sum up to 0 ($\sum_{p \in P} r_p(e) = 0$ holds for all edges $e \in E$), then we call the MMPG a zero-sum game.

The way that the respective players choose the successor vertex is a function $\sigma_p : V^*V_p \to V$ from an initial sequence of a play that ends in some vertex $v \in V_p$ of player $p$ to a vertex $v'$, such that $(v, v') \in E$. A family of strategies $\sigma = \{\sigma_p \mid p \in P\}$ is called a strategy profile. A strategy profile $\sigma$ defines a unique play $\pi_\sigma$, and therefore a reward $r_p(\sigma) = r_p(\pi_\sigma)$ for each player $p$.

A strategy profile is a Nash equilibrium if no player has an incentive to change her strategy, provided that all other player keep theirs. That is, for all players $p \in P$ and for all $\sigma' = \{\sigma'_q \mid q \in P\}$ with $\sigma_q = \sigma'_q$ for all $q \neq p$, $r_p(\sigma) \geq r_p(\sigma')$ holds.

A strategy profile is a *political equilibrium* for a designated player $d$ (for dictator), if no *other* player has an incentive to deviate her strategy. That is, for all players $p \in P \smallsetminus \{d\}$ and for all $\sigma' = \{\sigma'_q \mid q \in P\}$ with $\sigma_q = \sigma'_q$ for all $q \neq p$, $r_p(\sigma) \geq r_p(\sigma')$ holds. Thus, a political equilibrium allows for solutions, where the dictator could improve upon her reward by changing her strategy. While this may on first glance not be in the interest of a dictator, we will see that she can obtain better results with political than with Nash equilibria.

*Two player mean-payoff games* (2MPGs) can be viewed as a special case of multi-player mean-payoff games, where only two-players participate. 2MPGs are used in this paper to determine the outcome of MMPGs when, from some point onwards, one player, say $p$, is playing against all others, where the objective of $p$ is inherited from a multi-player MPG, while the objective of the remaining players is to minimise her reward. As the objective of the remaining players is defined by the objective of $p$, we use only $r_p$ to describe the objective of the game. The 2MPG *for* $p$ of an MMPG $\mathcal{M} = \langle P, V, \{V_p \mid p \in P\}, v_0, E, \{r_p : E \to \mathbb{Q} \mid p \in P\}\rangle$, denoted $\mathsf{2mpg}(\mathcal{M}, p)$, is thus the game $\langle P, V, \{V_p, V \smallsetminus V_p\}, v_0, E, r_p\rangle$. Mean-payoff games have optimal memoryless strategies for both players, and the outcome when starting in any vertex $v \in V$, is determined [24]. By abuse of notation, we denote this value by $r_p(v)$.

## 3 Optimal strategy profiles

We study the question of computing *optimal strategy profiles*. A strategy profile is called optimal if it is a Nash or political equilibrium and provides the maximal reward among the strategy profiles in this class of equilibria. In the remainder, we will show that

1. political equilibria are generally better than Nash equilibria (Theorem 1),
2. determining if there is a strategy profile $\sigma$ with $r_d(\sigma) = 1$, such that the strategy profile $\sigma$ is a Nash resp. political equilibrium, is NP hard even for zero-sum MMPGs with reward functions whose domain is in $\{-1, 1\}$ (Theorem 2), and optimal reward of the dictator is not efficiently approximisable (Corollary 1), and
3. optimal Nash and political equilibria always exist, and, for a fixed set of players, finding an optimal strategy profile in MMPGs is polynomial time reducable to solving 2 player MPGs (Corollary 3).

For social optima, it suffices to add a social reward to the reward function, e.g., the sum of the individual rewards, without letting the respective player own any vertex. The technique introduced in this paper can then be used to optimise the social payoff.

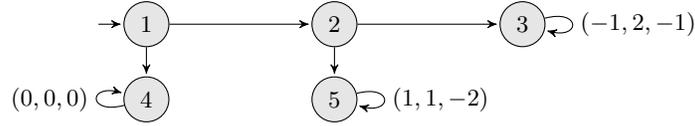

**Fig. 1.** An MMPG, where political equilibria are strictly better than Nash equilibria. The rewards are depicted in the order first player, dictator, passive player. The rewards on the edges $(1, 2)$, $(1, 4)$, $(2, 3)$, and $(2, 5)$ is not shown, because these edges can only be taken once in a play. Their rewards thus have no impact on the payoff for any player.

### 3.1 Superiority of political equilibria

In this subsection, we show that political equilibria is in general superior over Nash equilibria: a benign dictator who assigns strategies in such a way that she only makes sure that no *other* player has an incentive to deviate, while allowing for the use of 'modest' strategies that can be improved upon even if the other players stick to their strategies, is more successful than a dictator following the more obviously egoistic approach to chose only among strategies she cannot improve upon herself.

On first glance, it may not seem to be in the interest of the dictator to be benign. To the contrary, it would seem that the dictator could improve upon such strategy profiles by simply adjusting her strategy. A closer look, however, reveals that she only gets rid of constraints, and will therefore never deteriorate and often improve her reward.

To show this, consider the MMPG from Figure 3.1. It shows a simple MMPG with five vertices, 1 through 5, where the *dictator* owns Vertex 2, and a player *first* owns the initial vertex, Vertex 1. The other vertices have exactly one successor (themselves), such that it does not matter who owns them. There is a third player, player *passive*.

Initially, Player *first* can either play to Vertex 4, or to Vertex 2. When playing to Vertex 4, every player will receive a payoff of 0. When she plays to Vertex 2, the *dictator* can either move on to Vertex 3, securing herself a payoff of 2, to the cost of the *first* and the *passive* player, who both receive a payoff of $-1$. Alternatively, she could play to Vertex 5, where both the *dictator* and the *first* player receive a payoff of 1, to the cost of the *passive* player, whose payoff is $-2$.

In a Nash equilibrium, the *dictator* will never move to Vertex 5, as she can improve over such strategies by simply choosing to go to Vertex 3. Consequently, the *first* player will not move to Vertex 2 in any Nash equilibrium, as this would result in a payoff of $-1$ for her, such that moving to Vertex 4 is preferable. Thus, the only play produced by a Nash equilibrium is the play $1 \cdot 4^\omega$.

But the dictator has a better political equilibrium: she can benignly wave her option to move to Vertex 3, and instead move to Vertex 5. Then, it becomes preferable for the *first* player to move to Vertex 2. This results in an improved reward for the dictator.

**Theorem 1.** *Optimal political equilibria cannot be worse, but might be strictly better for the dictator compared to Nash equilibria.*

## 3.2 NP hardness

In order to establish NP hardness, we reduce the satisfiability of a 3SAT formula over $n$ atomic propositions with $m$ conjuncts to solving a zero-sum MMPG with $2n+1$ players and $4m+5n+2$ vertices that uses only payoffs 0 and 1. We consider the reduction for the example of the 3SAT formula $(p \vee \neg q \vee \neg r) \wedge (\neg p \vee q \vee \neg r) \wedge (\neg p \vee \neg q \vee \neg r)$.

The $2n+1$ players consists of $2n$ players for the $2n$ literals corresponding to the $n$ variables, and the *dictator* who intuitively tries to validate the formula. The game consists of three phases, an initial *assignment phase*, in which the *dictator* intuitively assigns either the value true or the value false to all $n$ variables. We use two-players for each of the variables, one representing *true*, and one representing *false*. In a second *validation phase*, the dictator intuitively validates that the chosen assignment indeed satisfies the specification $\varphi$. For this, she successively steps through the conjuncts of the 3SAT formula. For each conjunct, the dictator can select one of the three literals, which is owned by the same player who owns this literal in the first phase. In the first and second phase, the literal players can either continue, or move to an absorbing state. In a final *evaluation phase*, the game goes round a ring of length $n$, where, in each step, a disadvantage is given to one of the players of a variable, the player who represents true, or the player who represents false. By choosing the payoff for cycling in the absorbing state accordingly, we can assure that there is a political / Nash equilibrium with payoff 1 for the dictator if $\varphi$ is satisfiable, and a payoff of 0 for the dictator if $\varphi$ is unsatisfiable, using only the rewards 0 and 1.

**Assignment phase** In the assignment phase, we have two types of vertices. We have Vertices 0 through $n$, which are owned by the dictator and $2n$ literal vertices. In Vertex $i-1$, the dictator chooses either the truth value or false value of each variable $Z$ by either moving to a vertex $z_i$, or moving to $\neg z_i$, owned by the players with the respective literal. From vertex $z_i$ or $\neg z_i$, the respective player can choose to move to the dictator vertex, or to an absorbing vertex abs. From the absorbing vertex abs, there is just one outgoing transition, which returns to abs, and has a payoff of 0 for dictator and payoff of 1 for all other players. Note that payoff at the edges that can be taken only once are omitted as they have no effect on the overall reward of the play. The assignment phase is shown in the Figure 3.2.

**Validation phase** In the validation phase, the dictator intuitively tries to validate that her chosen assignment indeed validates the formula $\varphi$. Here, we have two types of vertices. We have $m$ dictator vertices and $3m$ literal vertices, $z_i^1$, $z_i^2$, and $z_i^3$ for all $i \in \{1, \ldots, m\}$. In this phase, the dictator successively steps through the conjuncts of the 3SAT formula. For each conjunct, the dictator selects one of the three literals, which is owned by the same player who owns this literal in the first phase. At vertex $n+i-1$, the dictator can play to any literal vertex, to $z_i^1$, to $z_i^2$, or to $z_i^3$ in conjunct $i$, to validate the value of the conjunct. Further, we have the same absorbing vertex as in the assignment phase. Here also, payoff at the edges that can be taken only once are omitted. The Validation phase is shown in the Figure 3.3.

**Evaluation phase** In evaluation phase, we have $2n$ literal vertices and a single dictator vertex. The evaluation phase of the MMPG resembles a ring structure. Here, the game cycles in a ring of length $n$ where at every vertex one of the players is at disadvantage. At any vertex $z_i$ resp. $\neg z_i$, its counter-literal receives a payoff of $0$ while everyone else receives payoff of $1$. The vertex owned by $z_i$ has two successors, the vertices owned by $z_{i\oplus 1}$ and $\neg z_{i\oplus 1}$, where $i \oplus 1$ is $i+i$ for $i \neq n$, and $n \oplus 1 = 1$. The vertex owned by $\neg z_i$ has the same successors as the vertex owned by $z_i$: the vertices owned by $z_{i\oplus 1}$ and $\neg z_{i\oplus 1}$. The Evaluation phase is shown in the Figure 3.4.

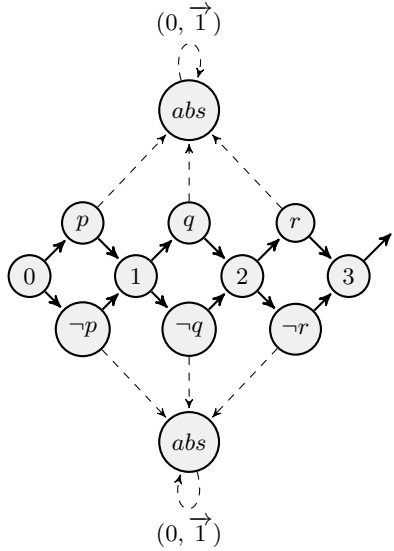
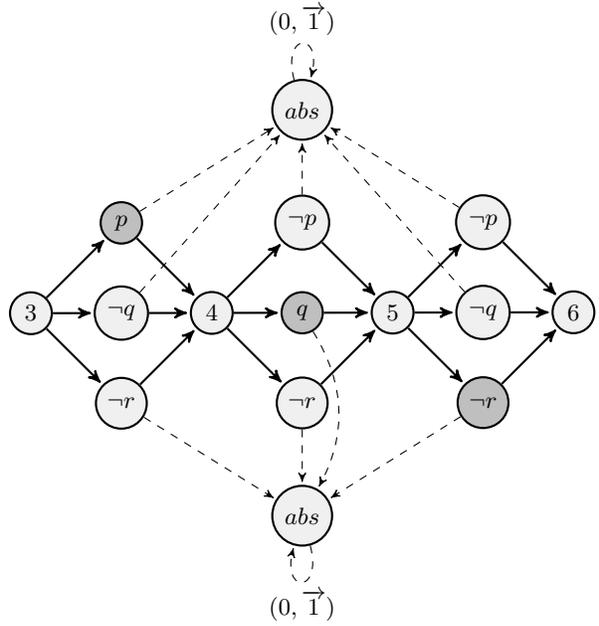

**Fig. 2.** Assignment phase      **Fig. 3.** Validation Phase

### 3.3 Approximisability

Note that, for *all* strategy profiles $\sigma$, the reward $r_d(\sigma)$ is either $0$ (if the absorbing state is reached) or $1$ (if the cycle in the evaluation phase is reached). In particular, the reward is $1$ if the 3SAT problem has a solution, and $0$ if it does not have a solution. Unless P equals NP, the optimal reward of the dictator therefore cannot be approximised closer than the trivial $0.5$ approximation by a tractable algorithm.

### 3.4 0 sum games

To progress from here to zero sum games, we can simply add a suitable number of additional players who own no vertex. If we maintain the rewards of $1$, replace the rewards of $0$ to $-1$, and assign a suitable number of these new players rewards of $-1$ and 1, respectively, such that the sum of the rewards is 0, we obtain a zero sum game, where the rewards of the dictator are either $-1$ or $1$ for all strategy profiles. The non-approximisability clearly carries over.

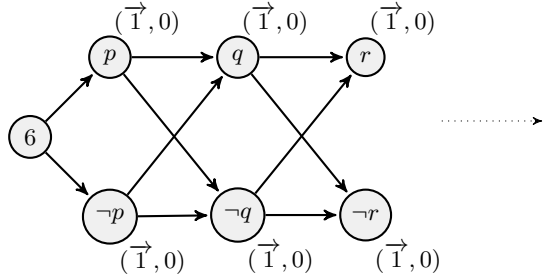

**Fig. 4.** Evaluation Phase

**Theorem 2.** *The decision problem of whether or not a political or Nash equilibrium $\sigma$ with reward $r_d(\sigma) = 1$ of the dictator exists in games with rewards in $\{0, 1\}$ resp. zero sum games with rewards in $\{-1, 1\}$, such that the reward of the dictator is always in $\{0, 1\}$ resp. $\{-1, 1\}$ is NP complete.*

The proof is closely related to the NP hardness proof from [23].

**Corollary 1.** *Unless P=NP, no tractable algorithm can approximate the optimal reward of the dictator closer than $0.5$.*

### 3.5 Reward and punish for political equilibria

Let us consider a strategy profile $\sigma$, which is a political equilibrium. We first show that we can obtain a similar equilibrium by applying a punishment to the first player who deviates from $\sigma$. The power to define the equilibrium allows the dictator to use the power of all remaining players to punish this deviator.

That is, we use a strategy profile where all players co-operate to produce $\pi_\sigma$. Note that the *dictator* solicits co-operation from every player who owns some vertex in the game. Further, the strategy profile $\sigma$ offers the reward $r_p(\pi_\sigma)$ to a player $p$, which is at least as good as the reward that player $p$ would have got in any Nash equilibria. But, if a player deviates from $\sigma$, all other players co-operate to harm this player, throwing their own interests to the wind. Thus, deviation from the construction of $\pi_\sigma$ will lead to a payoff of the deviating player, which equals the payoff of this player in a two-player game that starts at the point of her deviation, i.e., at the vertex owned by her where she is supposed to play as per $\sigma$. We call any such strategy profile a *reward and punish strategy profile* and define it as a $\sigma$ that offers reward $r_p(\pi_\sigma)$ to a player $p$ and any deviation from $\sigma$ by a player $p$ will eventually lead her to get a low payoff then $r_p(\pi_\sigma)$.

**Lemma 1.** *If a strategy profile $\sigma$ is a political equilibrium, then there is a reward and punish strategy profile $\sigma'$, which is also a political equilibrium and defines the same play $\pi_\sigma = \pi_{\sigma'}$. If $\sigma$ is Nash, so is $\sigma'$.*

*Proof.* First we observe that $\pi_\sigma$ alone defines the reward of all players for the strategy profile $\sigma$ and thus, due to $\pi_\sigma = \pi_{\sigma'}$, of $\sigma'$.

Let us assume for contradiction that a player $p \in P$ for Nash equilibria resp. $p \in P \smallsetminus \{d\}$ for political equilibria has an incentive to deviate from her strategy in $\sigma'$. Then her payoff in $\sigma'$ will be determined by the result of the two-player zero-sum MPG 'her against the rest' as defined by the reward and punish strategy profiles. (Note that the initial play up to this point has no impact on the limit reward.) But she can deviate from her strategy in $\sigma$ at the same position with at least the same reward, by simply assuming that she plays against all other players in the same game. Consequently, she has an incentive to deviate in the strategy profile $\sigma$, too, which contradicts the assumption that $\sigma$ is a Nash resp. political equilibrium. □

This observation allows us to concentrate on reward and punish strategy profiles only. Let $\text{ver}(\pi)$ be the set of vertices that occur in a play, and let $\text{own}(S) = \{p \in P \mid S \cap V_p \neq \emptyset\}$ be the set of players that own some vertex in $S$. With these terms, it is simple to characterise reward and punish strategy profiles.

**Lemma 2.** *For an MMPG $\mathcal{M}$, a play $\pi_\sigma$ is the outcome of a reward and punish strategy profile $\sigma$ which is a Nash resp. political equilibrium if, and only if, for all vertices $v \in \text{ver}(\pi)$ and all players $p \in \text{own}(\text{ver}(\pi_\sigma))$ resp. $p \in \text{own}(\text{ver}(\pi_\sigma)) \smallsetminus \{d\}$ that control a vertex that occurs in the play, it holds that $r_p(\pi_\sigma) \geq r_p(v)$.*

*Proof.* To show the 'if' direction, we assume for contradiction that $r_p(\pi_\sigma) < r_p(v)$ holds for some vertex $v \in \text{ver}(\pi)$, which is owned by $p$ (resp. owned by $p \neq d$). Then player $p$ can improve on her strategy by following her strategy until $v$ is reached, and henceforth follow the strategy from $2\text{mpg}(\mathcal{M}, p)$. As the initial play does not influence the limies inferior, her payoff would be at least $r_p(v)$, which is strictly greater than $r_p(\pi_\sigma)$. ↯

To show the 'only if' direction, we assume for contradiction that $r_p(\pi_\sigma) \geq r_p(v)$ holds, but no reward and punish strategy profile defines $\pi_\sigma$. Assume that Player $p$, deviates in vertex $v$ from $\pi_\sigma$. Then the other players will join to diminish her payoff henceforth. Taking into account that the initial sequence up to this point has no influence on the limit inferior of the payoffs, they can follow the optimal strategy of the oponents of $p$ from $2\text{mpg}(\mathcal{M}, p)$, restricting the payoff of player $p$ to $r_p(v)$. ↯ □

In the next step, we show that we can determine the existence of a *well behaved* reward and punish strategy profile that satisfies such a constraint system. A strategy profile is *well behaved* if the the ratio in which every edge occurs has a limit, that is, if, for all edges $(s,t) \in E$, there is a $r_{(s,t)} = \lim_{n \to \infty} \frac{\#_n^{(s,t)}(\pi_\sigma)}{n}$, where $\#_n^{(s,t)}(v_0, v_1, v_2 \ldots) = \left|\{i < n \mid (v_i, v_{i+1}) = (s,t)\}\right|$ is the number of edges $(s,t)$ among the first $n$ edges that occur in a play $v_0, v_1, v_2 \ldots$. (Recall that this limit does not necessarily exist for general strategy profiles.)

**Linear programs for well behaved reward and punish strategy profiles.** The first central observation is that if we already know

- the set of vertices $Q$ visited in $\pi_\sigma$ and
- a (strongly connected) set $S$ of vertices that are visited infinitely often,

then we can infer a constraint system by Lemma 2, which is necessary and sufficient for a well behaved reward and punish strategy profile. The constraint system consists of two parts. One part is the ratios, where we use the $p_{(s,t)}$ from above for edges $(s,t) \in E \cap S \times S$, and similarly $p_v$ for the limit ratio of each vertex in $S$. (Obviously, the limit ratio of each vertex not in $S$ and each edge not in $S \times S$ must be 0.)

This provides a first part of a constraint system, namely

- the ratio of vertices and edges that are not in $S$ resp. $S \times S$ is 0,
  $p_v = 0$ for all $v \in V \smallsetminus S$ and $p_e = 0$ for all $e \in E \smallsetminus S \times S$,
- the ratio of vertices and edges that are in $S$ resp. $S \times S$ is $\geq 0$,
  $p_v \geq 0$ for all $v \in S$ and $p_e \geq 0$ for all $e \in E \cap S \times S$,
- the sum of the ratio of vertices is 1, $\sum_{v \in V} p_v = 1$, and
- the ratio of a vertex is the sum of the ratios of its incoming and outgoing edges,
  $p_s = \sum_{t.(s,t) \in E} p_{(s,t)}$ for all $s \in S$ and $p_t = \sum_{s.(s,t) \in E} p_{(s,t)}$ for all $t \in S$.

The second part of the constraint system stems from Lemma 2: as $r_p(\pi_\sigma)$ is simply $\sum_{e \in E} p_e r_p(e)$, that is, it is the weighted sum of the rewards of the individual edges, we get the constraints

$$\sum_{e \in E} p_e r_p(e) \geq \max_{v \in Q}(r_p(v))$$

for all $p \in \mathsf{own}(Q)$ for Nash, and for all $p \in \mathsf{own}(Q) \smallsetminus \{d\}$ for political equilibria.

Before we define the objective function, we state a simple corollary from Lemma 2.

**Corollary 2.** *Every well behaved reward and punish strategy profile satisfies these constraints, and every well behaved strategy profile $\sigma$, whose play $\pi_\sigma$ satisfies these constraints, defines a reward and punish strategy profile.* □

The objective of the dictator is obviously to maximise $r_d(\pi_\sigma) = \sum_{e \in E} p_e r_d(e)$. Once we have this linear programming problem, it is simple to determine a solution in polynomial time [12, 13].

The relevant points are first to establish that a well behaved reward and punish strategy profile exists for each such solutions, and second to show that non-well behaved reward and punish strategy profiles cannot be preferable for the dictator.

*From $Q$, $S$, and a solution to the linear programs to a well behaved reward and punish strategy profile.* We start with the simple case that the vertices and edges with non-0 ratio are strongly connected.

We design $\pi_\sigma$ as follows. We first go from the initial vertex $v_0$ through states in $Q$ to some state in $S$. (Note that this initial path has no bearing on the limes inferior that defines the payoff of the individual players.)

Once we have reached $S$, we intuitive keep a list for each vertex in $S$. In this list, we keep the number of times each outgoing edges with non-0 ratio has been taken. We also apply an arbitrary (but fixed) order on the outgoing edges. Each time we are in this vertex, we choose the first edge (according to this order) that has been taken less often (from this vertex) than $\frac{p_e}{p_v}$, the ratio $p_e$ of the edge divided by the ratio $p_v$ of this vertex, suggests. If no such edge exists, we take the first edge.

The result is obviously a well behaved strategy profile and the first part of the constraint system is clearly satisfied. It therefore suffices to convince ourselves that the second part is satisfied as well.

Now assume for contradiction that this is not the case. Let $q_v$ and $q_e$ be the real ratio of the vertices and edges, respectively. Note that our simple rule for the selection of vertices implies that $\frac{p_e}{p_v}$ is correct for all edges $e = (v, v') \in E \cap S \times S$. Then there must be a vertex $v \in S$, which has the highest factor $\frac{q_v}{p_v}$. As it is the highest factor, none of its predecessors in $E \cap S \times S$ can have a higher ratio; consequently, they must have the same ratio. By a simple inductive argument, this expands to the complete strongly connected set of non-0 vertices. As $\sum_{v \in S} p_v = 1 = \sum_{v \in S} q_v$ holds, this implies $p_v = q_v$ for all $v \in S$.

To extend this argument to the general case, we first observe that the non-0 vertices and edges form islands of (maximal) SC parts $C_1$, through $C_k$. We use this observation to compose a play as follows.

We start with an initial part, a transfer from $v_0$ to $C_1$ as in the simple case. We then continue by playing a $C_1^1$ part, a transfer, a $C_2^1$ part, a transfer, ..., a $C_k^1$ part, transfer $C_1^2$, and so forth. To achieve a well behaved strategy profile we do the following.

1. We fix the ratio $\sum_i C_1^i : \sum_i C_2^i : \ldots : \sum_i C_k^i$ according to the the sum of the $p_v$ for vertices $v$ in the respective component. This ratio never changes, and it is given by natural numbers $c_1, c_2, \ldots, c_k$, such that $c_1 : c_2 : \ldots : c_k$ satisfies this ratio.
2. We let $C_j^i$ grow slowly with $i$. We can, for example, use $i \cdot c_j$.
   Note that the transfer part has constant length, bounded by $|S|$. Thus the limit ratio of transfer is 0.
3. We let the transfer to $C_{j+1}^i$ go to the vertex, in which $C_j^i$ was left. Note that the transfer may contain vertices of various components, but as the overall ratio of the transport is 0, this does not affect the limit probability.

   Thus, we can use the controller from the simple case of one SCC for the sequence $C_i^1, C_i^2, C_i^3 \ldots$, which only focuses on the relevant part of the $i^{th}$ component.

In effect, we have simple controllers for the individual components, and a single counting controller that manages the transfer between the components.

It is easy to see that the resulting controller inherits the right ratios from the simple individual controllers. Together with Corollary 2 we get:

**Theorem 3.** *If the linear program from above for sets $Q$ of reachable states and $S$ of states visited infinitely often has a solution, then there is a well behaved reward and punish strategy profile that meets this solution.* □

Finally, we show that non-well behaved reward and punish strategy profiles cannot provide a better solution than the one provided by the previous theorem.

**Theorem 4.** *Non-well behaved reward and punish strategy profiles cannot provide better rewards for the dictator than the reward $r_d$ for the dictator obtained by the well behaved reward and punish strategy profiles described above.*

*Proof.* We have shown in Lemma 2 that there exists a well defined constraint system obeyed by all reward and punish strategy profiles with set $Q$ of reachable states and all $p \in \mathsf{own}(Q)$ for Nash, and for all $p \in \mathsf{own}(Q) \smallsetminus \{d\}$ for political equilibria. Let us assume for contradiction that there is a reward and punish strategy profile $\sigma$ that defines a play $\pi_\sigma$ with a strictly better reward $r_d(\pi_\sigma) = r_d + \varepsilon$ for some $\varepsilon > 0$.

Let $k$ be some position in $\pi_\sigma$ such that, for all $i \geq k$, only positions in the infinity set $S$ of $\pi_\sigma$ occur. Let $\pi$ be the tail $v_k v_{k+1} v_{k+2} \ldots$ of $\pi_\sigma$ that starts in position $k$. Obviously $r_p(\pi) = r_p(\pi_\sigma)$ holds for all players $p \in P$.

We observe that, for all $\delta > 0$, there is an $l \in \mathbb{N}$ such that, for all $m \geq l$, $\frac{1}{m}\sum_{i=0}^{m-1} r_p\bigl((v_i, v_{i+1})\bigr) > r_p(\pi) - \delta$ holds for all $p \in P$, as otherwise the limes inferior property would be violated.

We now fix, for all $a \in \mathbb{N}$, a sequence $\pi_a = v_k v_{k+1} v_{k+2} \ldots v_{k+m_a}$, such that $v_{k+m_a+1} = v_k$ and $\frac{1}{m}\sum_{i=0}^{m_a-1} r_p\bigl((v_i, v_{i+1})\bigr) > r_p(\pi) - \frac{1}{a}$ holds for all $p \in P$.

Let $\pi_0 = v_0 v_1 \ldots v_{k-1}$. We now select $\pi' = \pi_0 \pi_1^{b_1} \pi_2^{b_2} \pi_3^{b_3} \ldots$, where the $b_i$ are natural numbers big enough to guarantee that $\frac{b_i \cdot |\pi_i|}{|\pi_{i+1}| + |\pi_0| + \sum_{j=1}^{i} b_j \cdot |\pi_j|} \geq 1 - \frac{1}{i}$ holds.

Letting $b_i$ grow this fast ensures that the payoff, which is at least $r_p(\pi) - \frac{1}{i}$ for all players $p \in P$, dominates till the end of the first iteration[1] of $|\pi_{i+1}|$.

The resulting play belongs to a well behaved (as the limit exists) strategy profile, and can thus be obtained by a well behaved reward and punish strategy profile by Lemma 2. It thus provides a solution to the linear program from above, which contradicts our assumption. □

**Decision & optimisation procedures.** The *decision problem* related to the construction of optimal equilibria asks whether or not, for a given threshold $r_{\mathsf{thld}}$, there exists a strategy profile $\sigma$, which is a Nash resp. political equilibrium and provides a reward $r_d(\pi_\sigma) \geq r_{\mathsf{thld}}$ for the dictator.

In Lemma 2 and Theorem 4 we have established that it is enough to consider well behaved reward and punish strategy profiles. The relevant behaviour of these strategy profiles is captured by the set of reachable vertices, the set of infinite vertices $S$, and the ratio of the edges in $E \cap S \times S$.

We use this observation in various algorithms, starting with a nondeterministic one.

**Theorem 5.** *For an MMPG $\mathcal{M}$ and a threshold $r_{\mathsf{thld}}$, the respective decision problem for political or Nash equilibria is NP complete, both in the general case and when restricted to zero-sum games with pay offs in $\{-1, 1\}$.*

*Proof.* We use nondeterminism to guess a set $Q$ of visited vertices, a set $S$ of vertices visited infinitely often and then the linear program defined by them and a solution thereof. Note that the linear program is polynomial in $\mathcal{M}$ and, consequently, it has a polynomial solution, too.

---

[1] Including the first iteration of $\pi_{i+1}$ is a technical necessity, as a complete iteration of $\pi_{i+i}$ provides better guarantees, but without the inclusion of this guarantee, the $\pi_j$'s might grow too fast, preventing the existence of a limes.

After having a closer look at the sets $Q$ and $S$, we can check that there is a possible path from the initial vertex to $S$, that $S$ is strongly connected, that $Q$ and $S$ define the guessed linear program, its constraint system is satisfied by the solution and the reward of the dictator is at least the threshold $r_{\mathsf{thld}}$ given. All of these tests can obviously be performed in polynomial time.

The respective hardness results have been established in Theorem 6.

Although there is no perfectly fitting lemma or theorem for citing in, the inclusion in NP could have been cited in from [23] and Theorem 5, and the techniques used there are quite similar to ours. We re-proved it as we need the intermediate results below.

The hardness result uses a polynomial number of players. This raises a question if the complexity is better for a bounded number of up to $k$ players.

We first assume that we are already provided with solutions to the 2MPGs to $\mathcal{M}$. To device a decision procedure, we start with a simple observation:

**Lemma 3.** *For a given MMPG $\mathcal{M}$ with $k$ players and $n$ vertices, there are at most $(n+1)^k$ many different thresholds in the related linear programs.*

*Proof.* For each player $p$, there is either the threshold $r_p(v)$ for some vertex $v$ of $\mathcal{M}$, or no restriction on the threshold at all in Part II of the constraint system of a linear program. □

Consequently, we only have to consider the most liberal constraint systems.

**Lemma 4.** *For a given MMPG $\mathcal{M}$ with $k$ players and $n$ vertices and a threshold as referred to in the proof of Lemma 3, it suffices to refer to up to $n$ first parts of the constraint system of the LLP.*

*Proof.* For each Part II of the constraint system as referred to in the proof of Lemma 3, it is easy to determine the maximal set $Q$ of nodes that can be visited. For this maximal $Q$, we can determine the strongly connected components $S_1, S_2, \ldots$ of $(V, E) \cap Q$ that are reachable from the initial vertex $v_0$. Obviously, there are at most $n$ of them.

It is now easy to see that, for all $Q'$, $S'$ that define Part II of the constraint system, $Q'$ is contained in $Q$ and $S'$ is contained in one SCC $S_i$ from above. Now $Q$ and $S_i$ define a more liberal Part I of a constraint system than $Q'$ and $S'$. Thus, every solution for $Q'$ and $S'$ is a solution for $Q$ and $S_i$, too. □

Thus, for a given $k$, there are only polynomially many LLPs to consider, and they are easy to construct. Solving LLPs requires only polynomial time [13, 12]. We thus get:

**Theorem 6.** *If we are provided with the solutions to the 2MPGs defined by an MMPG with a fixed number $k$ of players, then we can determine an optimal solution in polynomial time.* □

**Corollary 3.** *MMPGs with a fixed number of players can be solved in polynomial time by a machine with an oracle for solving two-player zero-sum MPGs. If 2MPGs are solvable in polynomial time, so are MMPGs with a fixed number of players.* □

### 3.6 Reduction to Two Player Mean-pay off games

Thus, finding optimal strategy profiles in MMPGs with a fixed number of players can be derived from solutions to 2MPGs. Various works have been published on solving 2MPGs. In [6], the authors give an improved pseudopolynomial procedure to solve two-player mean-payoff games. [2] provides a randomised strongly subexponential and pseudopolynomial algorithm, and [11, 24] contain an UP∩CoUP reduction. There are wilder reductions like one to symbolic linear programming [20] and a smoothed polynomial time complexity [3].

Corollary 3 therefore provides:

**Corollary 4.** *MMPGs with a fixed number of players can be solved in UP∩coUP, in pseudo polynomial time, in smoothed polynomial time, and in randomised subexponential time.* □

### 3.7 Making Rules in Good Cause

While the mechanism described refers to a selfish dictator, we would like to point out that the same mechanism can be used for finding socially optimal equilibria and, more generally, equilibria optimal for any ordered vector of payoffs.

For social optima, all one needs to do is to add a social reward to the reward function—traditionally the sum of the individual rewards—without letting the respective player own any vertex. The technique introduced in this paper can then be used to optimise the social payoff. Likewise, a dictator might choose to optimise her payoff first, but take a social optimum as a secondary objective. In this case, one would simply use the techniques discussed earlier to determine her optimal payoff, and then add this payoff as a constraint in the second part of the constraint system. Subsequently, one would repeat the process with the objective to optimise the social payoff. (Obviously, a more friendly dictator might choose to reverse the order of priorities.)

This 'optimise – add to constraint system – optimise' technique can obviously be generalised to an arbitrary number of objectives.

## 4 Discussion

The two main contributions of this paper are the introduction of political equilibria and the concept of well behaved reward and punish strategy profiles. Well behaved reward and punish strategy profiles are general instruments for optimising the payoff of one player, while projecting away problems like the potential non-existence of limit average values. It is our believe that they will be useful in many related optimisation problems. The introduction of political equilibria is a conceptual change of Nash equilibria, where an interested party overcomes an antinomy of Nash equilibria exemplified in Figure 3.1: the interested party (which we christianed the dictator) might improve her payoff by choosing a strategy, which is not stable for herself in the Nash sense of not being able to improve the payoff by unilaterally deviating from her strategy.

The solutions one obtains can be used to make stable rules that optimise various outcomes, including social optima as well as egoistic solutions.